# Superconducting phases of YH$_9$ under pressure

Mingyang Du[1], Zonglun Li[1], Defang Duan[1,*], Tian Cui[2,1,*]

[1] State Key Laboratory of Superhard Materials, College of Physics, Jilin University, Changchun 130012, People's Republic of China

[2] Institute of High Pressure Physics, School of Physical Science and Technology, Ningbo University, Ningbo, 315211, People's Republic of China

KEYWORDS: Hydrides, High pressure, Superconductivity, First principles calculation.

*Correspondence authors. duandf@jlu.edu.cn, cuitian@nbu.edu.cn

**ABSTRACT**

Yttrium superhydrides has attracted much attention due to their multiple stoichiometries and excellent superconductivities under high pressure. Especially, YH$_9$ have a $T_c$ of 243 K, which is only second to LaH$_{10}$ among all superconductors. It exhibits a positive pressure dependence of $T_c$ below 200 GPa, contrary to the results of theoretical prediction. In order to explore the origin of $T_c$ at low pressure, we extensively investigated the crystal structure of YH$_9$ at different pressure, and found a distorted cage structure with a symmetry of *Pnma*. This phase has the lowest enthalpy at pressure below 220 GPa, and its X-ray diffraction patterns is consistent with experimental data. Most importantly, the pressure dependence of $T_c$ in phase *Pnma* is in line with the experimental results. Further calculations show the structural distortion strongly affect the lattice vibration and electron-phonon coupling, then the *Pnma* phase exhibiting a positive pressure dependence of $T_c$.

**Introduction**

As early as in 1935, E. Winger and H. B. Huntington predicted that insulated molecular hydrogen transformed into metal atomic under high pressure[1]. And the metallic hydrogen was predicted to be an ideal high-temperature or even room-temperature superconductor[2]. However, hydrogen metallization was not clearly observed in the experiment even under the extreme high pressure close to 500 GPa[3,4]. In order to realize hydrogen metallization within the pressure range that diamond anvil cell (DAC) can reach, an alternative scheme is proposed — superhydrides.

Introducing other elements into hydrogen produce a "chemical precompression" effect on the hydrogen sublattice, so that hydrogen can exist as a stable atomic phase at a much lower pressure in superhydrides[5]. Inspired by this, many superhydrides have been designed that are expected to exhibit high $T_c$s[6-8], some of which have been experimentally confirmed. In particular, $H_3S$ and $LaH_{10}$ have high $T_c$s exceeding 200 K by theoretical predicted and experimental measurement [9-11] [12-15].

Among these hydrogen-based superconductors, clathrate superhydrides have attracted extensive attention due to their outstanding superconductivity. Such structures are common in alkaline earth metals superhydrides and rare earth metals superhydrides $EH_n$ (n = 6, 9, 10), such as binary hydrides $CaH_6$[16], $MgH_6$[17], $YH_{6,9,10}$[10,11,18], $ScH_6$[19], $(Tm/Yb/Lu)H_6$[20], and ternary hydrides $(Y,Ca)H_6$[21-23], $(Mg,Ca)H_6$[24], $(Sc,Ca)H_6$[25], $(La,Y)H_6$[26], $(Ca/Sc/Y,Yb/Lu)H_6$[27]. These superhydrides contain $H_{24}$, $H_{29}$ or $H_{32}$ cages, and metal elements that produce chemical pre-compression to maintain the stability of the hydrogen cage. After $LaH_{10}$ confirmed the great potential of clathrate superhydrides in high-temperature superconductivity, more clathrate superhydrides were experimentally synthesized, including ThH$_9$[28], ThH$_{10}$[28], YH$_9$[29-31], CaH$_6$[32,33], YH$_6$[34], CeH$_9$[35], (La,Y)H$_{10}$[36], (La,Ce)H$_{9,10}$[37,38].

Yttrium superhydrides have been attracting numerous studies due to their excellent properties and multiple stoichiometry. Yttrium superhydrides with the composition YH$_3$, YH$_4$, YH$_6$ and YH$_9$ have been synthesized. Among them, YH$_4$ (88 K at 155 GPa[39]), YH$_6$ (224 K at 166 GPa[34]), YH$_9$ (243 K at 201 GPa[29], 262 K at 182 GPa[30], 230 K at 300 GPa[31]) exhibit superconductivity. The theoretical predicted superconductivity of YH$_3$ (40 K at 17.7 GPa) [40], YH$_7$ (21.5-43 K at 165 GPa[34]) and YH$_{10}$ (305-326 K at 250 GPa[10]) needs to be further verified experimentally.

YH$_9$ has the highest $T_c$ among the synthesized yttrium superhydrides, and its $T_c$ is second only to LaH$_{10}$ (250 K at 170 GPa[14], 260 K at 180 GPa[15]). At first, $T_c$ increases rapidly with the increase of pressure and reaches a maximum value of 243 K at around 200 GPa, and then $T_c$ decreases with the increase of pressure. This complicated pressure dependence of $T_c$ may be associated with the structural distortions and phase transformations. It is reported that some hydrogen-based superconductors are observed to be stable in phases with high symmetry and high $T_c$ under high pressure, and in phases with low symmetry and low $T_c$ under low pressure, such as H$_3$S[41] and LaH$_{10}$[42]. For YH$_9$, the *P*6$_3$/*mmc* phase was experimentally confirmed to be the high symmetry and high $T_c$ phase above 200 GPa. For the lower $T_c$ below 200 GPa, it maybe come from a lower symmetry structure. In recent years, several structures of YH$_9$ have been predicted, such as *P*6$_3$/*m*[11], *Cmcm*[11], *P*1[34] and *F*-43*m*[34,43]. But their stable pressure range, pressure dependence of $T_c$ or X-ray diffraction (XRD) are not in good agreement with experimental results. Therefore, the crystal structure and superconductivity of YH$_9$ below 200 GPa deserve to be further explored.

In this work, we extensively investigated the high-pressure phases of YH$_9$, and uncovered a hitherto unknown phase with symmetry of *Pnma* between 157 and 220 GPa. The X-ray diffraction patterns of *Pnma* phase is consistent with previous experimental data. In addition, the $T_c$ of *Pnma* phase increases rapidly with compression which is also in line with the experimental results. Therefore, the low-$T_c$ phase of YH$_9$ is most probably from the *Pnma* phase.

**Computational details**

The candidate crystal structure of YH$_9$ were explored by Ab Initio Random Structure Searching technique[44,45] with Cambridge Serial Total Energy Package [46]. Preliminary structural relaxation are performed using the Perdew-Burke-Ernzerhof parametrization of the generalized gradient approximation[47] for the exchange-correlation functional and on-the-fly generation of ultra-soft potential.

Final structural relaxation, enthalpies calculations and all electronic properties are calculated by Vienna Ab initio Simulation Program[48]. Brillouin zone (BZ) sampling using Monkhorst-Pack[49] meshes with resolutions of 2π×0.02 Å$^{-1}$. The projector augmented plane-wave potentials[50] set cut-off energy of 1000 eV. All parameters were selected to ensure that enthalpy calculations are well converged to be better than 1 meV per atom.

We use Quantum-ESPRESSO[51] for electron−phonon coupling (EPC) calculations. Ultra-soft potentials were used with suitable cut-off energy of 90 Ry. The k-points and q-points meshes in the first BZ are 24×24×24 and 6×6×6 for YH$_6$, 12×12×6 and 6×6×3 for *Cmcm* phase, 12×12×12 and 3×3×3 for *Pnma* phase, 12×12×9 and 6×6×3 for *P6₃/mmc* phase. $T_c$ of these structures are calculated by the Allen−Dynes-modified McMillan equation (A-D-M) with correction factors[52,53] and self-consistent iteration solution of the Eliashberg equation (scE)[54].

**Results and discussion**

Ab initio random structure searches were focused on YH$_9$ with 1 to 4 formula units under pressure of 100-300 GPa. A new stable phase with symmetry of *Pnma* (4 f.u./cell) were uncovered in the pressure range considered. Their crystal structures are listed in Table S1 of the supplementary information.

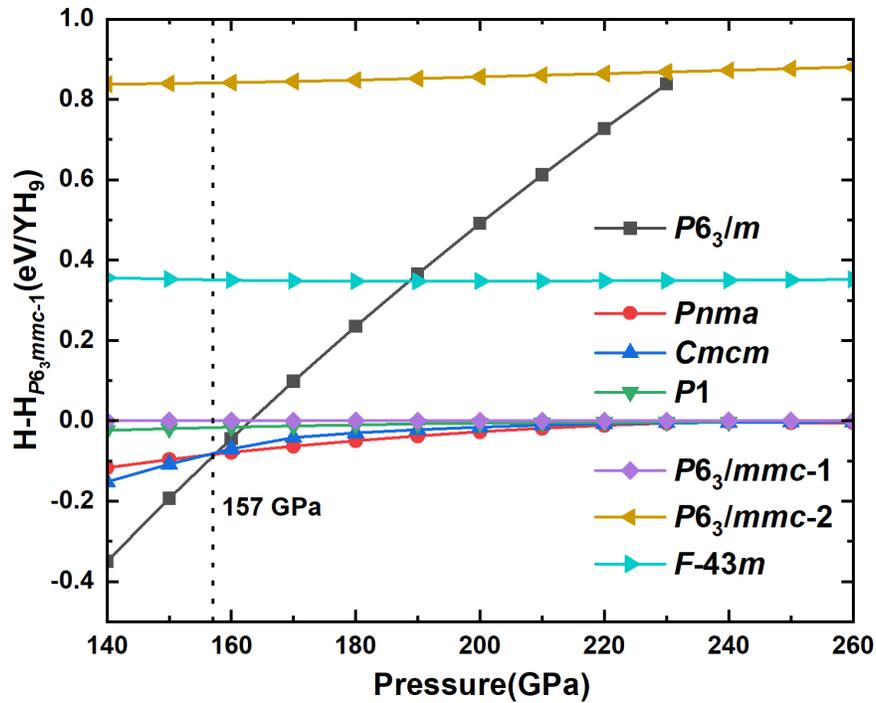

Fig. 1. Calculated enthalpies of various phases relative to the *P6₃/mmc*-1 phase as a function of pressure.

Fig. 1 shows the pressure dependence of enthalpies for all phases that are found in this work and previous studies. *P6₃/m* phase is the most stable phase at pressure below 157 GPa, which is a layered structure, and each layer is composed of Y atom and $H_3$ molecular units (see Fig. 2a). The enthalpy of the *Pnma* phase is lower than that of the $P6_3/m$ phase at the pressure above 157 GPa, and *Pnma* phase becomes the most stable one. Finally, the highly symmetric $P6_3/mmc$ structure becomes favorable above 220 GPa. It is worth noting that, there are two different phases with $P6_3/mmc$ space group. We denote the phase proposed by Peng et al.[11] as $P6_3/mmc$-1 and the phase proposed by Troyan et al.[34] as $P6_3/mmc$-2. The $P6_3/mmc$-1 phase is a kind of clathrate structure with $H_{29}$ cage, which is widely exists in rare earth metal superhydrides at high pressure. In this structure, H atoms are bonded covalently to to form hydrogen cage, and rare earth metal atoms are located in the centers of the hydrogen cages (see Fig. 2b). The crystal structure of *Pnma*, *Cmcm* and *P*1 phases are all similar to that of $P6_3/mmc$-1 phase (see Fig. 2c-e), except that their $H_{29}$ cage has different degrees of distortion. With the increase of pressure, the degree of distortion is narrowing, and they transform into the $P6_3/mmc$-1 phase above 220 GPa. The *F*-43*m* and $P6_3/mmc$-2 are metastable phases with high enthalpies. The *F*-43*m* phase contains $H_{28}$ cage, equivalent to a twisted $H_{32}$ cage (such as $LaH_{10}$[10]) missing four H atoms (see Fig. 2f), and similar structure exists in $PrH_9$[55]. The $P6_3/mmc$-2 phase has a spindle type $H_{29}$ cage which is different with the $P6_3/mmc$-1 phase (see Fig. 2f).

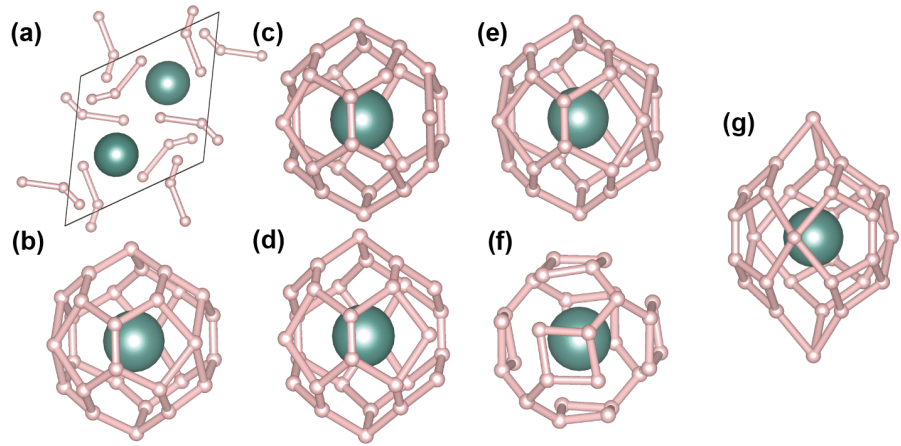

Fig. 2. Crystal structures of (a) $P6_3/m$, (b) $P6_3/mmc$-1, (c) *Pnma*, (d) *Cmcm*, (e) *P*1, (f) *F*-43*m* and (g) $P6_3/mmc$-2.

To determine the dynamic stability of $P6_3/m$, *Pnma*, *Cmcm*, and $P6_3/mmc$-1 phases, we calculated their phonon spectrum at different pressures. For $P6_3/m$ phase, lack of imaginary frequency modes at 140 and 150 GPa (see Fig S1) indicated its dynamically stability in this pressure range. *Cmcm* phase can be dynamically stable between 100 and 150 GPa (see Fig S2), while *Pnma* phase is dynamically stable in the pressure range of 160-210 GPa (see Fig S3). This is consistent with the thermodynamically stability that *Pnma* phase has lower enthalpy than *Cmcm* above 157 GPa. As the case of $P6_3/mmc$-1 phase, it can be dynamically stable above 250 GPa (see Fig S4), while emergence of imaginary frequency modes below 250 GPa indicates its dynamic instability. There would be an uncovered stable structure with more atoms, or the anharmonic effects may cause the dynamic stabilities of $P6_3/mmc$-1 phase between 210 and 250 GPa. However, the detailed study of this requires a rather large amount of computation, which we will probably carry out in future work.

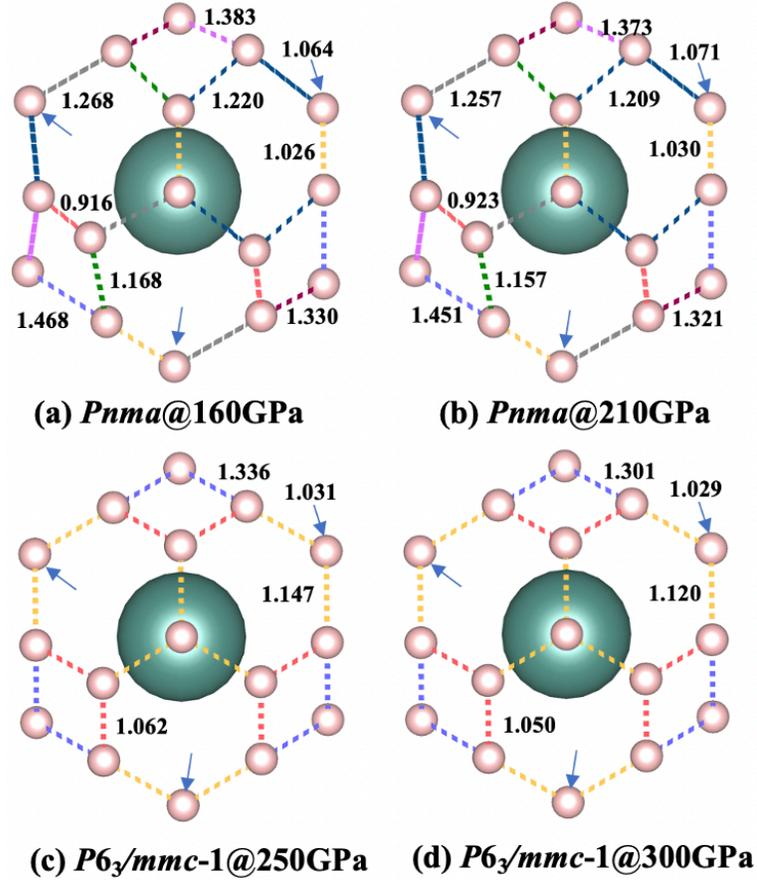

Fig. 3. H-H distances (Å) in $H_{29}$ cage of *Pnma* phase at (a) 160 GPa and (b) 210 GPa, $P6_3/mmc$-1 phase at (c) 250 GPa and (d) 300 GPa. The H-H bond with the same color in the same structure have the same length. The three H atoms pointed by the arrow have H-H bonds perpendicular to the plane, whose lengths are marked on the arrows at the upper right corner.

To compare the structural distortion of $H_{29}$ cage in *Pnma* phase with that in $P6_3/mmc$-1 phase, their detailed configurations of $H_{29}$ cage at different pressures and the corresponding H-H distances are shown in Fig. 3. *Pnma* phase has nine kinds of H-H bonds with different lengths induce the distortion of the $H_{29}$ cages. As the pressure increases from 160 GPa to 210 GPa, all H-H bonds longer than 1.1 Å shorten, while three kinds of H-H bonds shorter than 1.1 Å (marked with red, yellow and arrow) stretch (see Fig. 3a and b). $P6_3/mmc$-1 phase has only four kinds of H-H bonds and all H-H bonds shorten with increasing pressure (see Fig. 3c and d).

We furthermore calculated the Bader charge of *Pnma* and *P6₃/mmc*-1 phases at different pressures (see Table S2). In the *Pnma* phase at 160 GPa, the charges accepted per H atom varies from 0.1 to 0.35 e. Five different kinds of H atoms can be distinguished according to the charges obtained. As the pressure rises to 210 GPa, the electrons gained by H atoms more than 0.14 e decrease, while that less than 0.14 e increase. *P6₃/mmc*-1 phase with the high symmetry has three kinds of H atoms and the electrons gained by H atoms reduce with the increase of pressure. The electrons obtained by H atoms can fill the antibonding state of the H-H bonds, which affects the length of H-H bond. Therefore, we can see that the essential reason why the H-H bond length converges with the increase of pressure is that the pressure make the charge more evenly distributed, which is actually conducive to superconductivity. For *P6₃/mmc*-1 phase, the electrons accepted per H atom donated by the Y metal is basically evenly distributed. However, as the pressure increases, the total charges transferred by the Y atom decreases, which leads to a decrease in charges gained by all H atoms, thus leading to the shortening of H-H bonds.

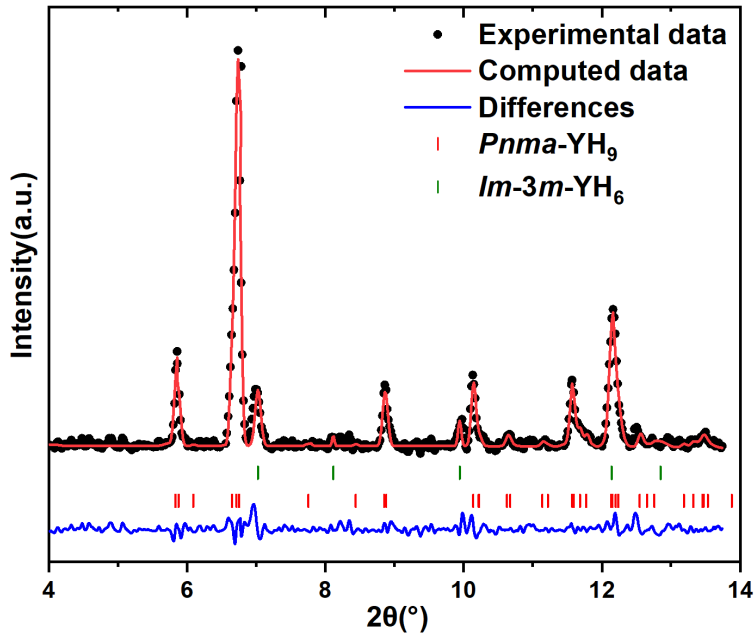

Fig. 4. X-ray diffraction patterns of the synthesized yttrium hydrides by Kong et. al. [29] compared with predicted *Pnma*-YH$_9$ and *Im-3m*-YH$_6$ at 255 GPa.

Due to the lower enthalpy and similar structure, we consider that the previously synthesized yttrium hydrides also contain *Pnma* phase, so we calculated the X-ray diffraction patterns (XRD) of *Pnma* phase and compared it with the experimental data measured by Kong et al.[29] (see Fig. 2). The results show that the XRD of *Pnma* phase is consistent with that of experimentally synthesized yttrium hydrides, except for two peaks corresponding to YH$_6$ (at 2θ=7 and 10° in Fig. 2). This indicates that the *Pnma* phase is very likely to be the phase with low symmetry and low $T_c$ we are looking for.

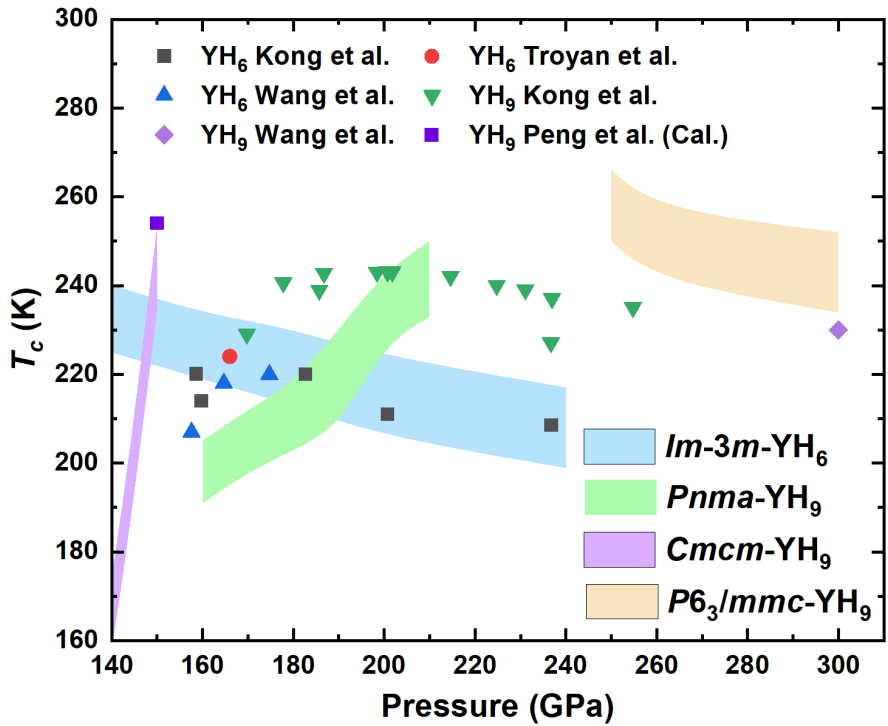

Fig. 5. Pressure dependence of $T_c$ for *Im-3m* phase of YH$_6$, and *Cmcm*, *Pnma*, *P6$_3$/mmc*-1 phases of YH$_9$. The stained areas represent the results of theoretical prediction, and the points represent the results of previous theoretical prediction[11] and experimental measurement[29,31,34].

Fig. 5 shows the pressure dependence of critical temperature for *Im-3m* phase of YH$_6$, and *Cmcm*, *Pnma*, *P6$_3$/mmc*-1 phases of YH$_9$. The stained areas represent the results estimated through the Eliashberg equation with Coulomb pseudopotential of μ* = 0.1 and 0.13. The points represent the result of theoretical prediction by Peng et. al.[11], experimental observations by Kong et al.[29], Troyan et al.[34] and Wang et al.[31]. Our prediction of YH$_6$ (blue area) is good agreement with the experimental observation (black points), which indicates that our calculations are accurate for yttrium hydrides. It can be seen from Fig. 5 that the estimated $T_c$s of *Cmcm* phase are 152-167 K at 140 GPa, and 235-253 K at 150 GPa, which deviate from experimental data. Note that the estimated $T_c$s of *Pnma* increase with increasing pressure, from 191-205 K at 160 GPa to 233-250 K at 210 GPa, which is in line with the pressure dependence of $T_c$ in experiments. Above 250 GPa, the predicted $T_c$s of *P*6$_3$/*mmc*-1 phase, decrease with increasing pressure (250-266 K at 250 GPa to 234-252 K at 300 GPa), which is consistent with the experimental trend.

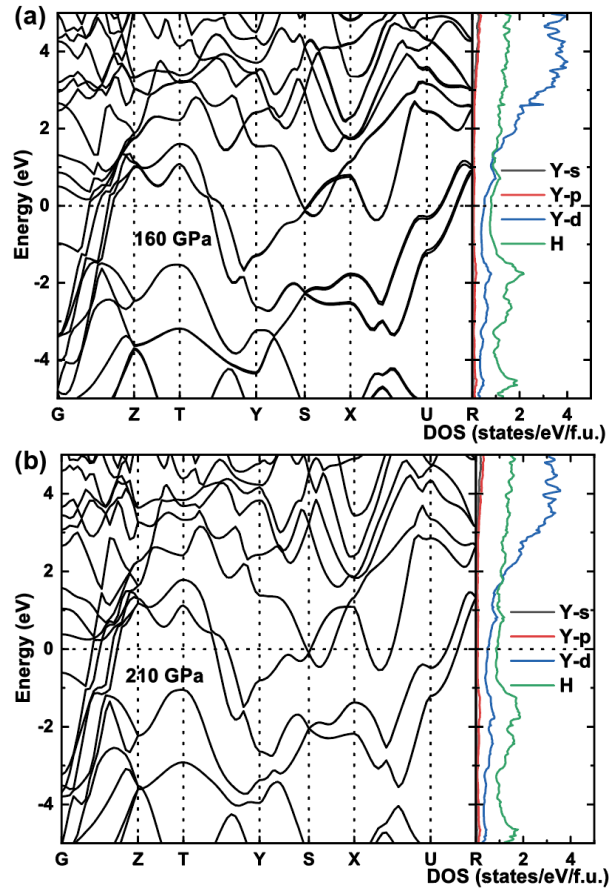

Fig. 6. Calculated electronic band structures and projected density of states (PDOS) for *Pnma* phase at (a) 160 GPa and (b) 210 GPa.

Subsequently, we calculated the electronic band structures and projected density of states (PDOS) to understand the effect of pressure on the electronic structure for the *Pnma* phase. As shown in Fig. 6, contributions from H s-orbitals and Y d-orbitals dominate the DOS at the Fermi level. At 160 GPa, the energy band structure of the *Pnma* phase has many split energy bands along the X-U-R path (see Fig. 6a). When the pressure rises to 210 GPa, these bands degenerate (see Fig. 6b). It may be related to the diminish of $H_{29}$ cage distortion in *Pnma* phase with the increase of pressure.

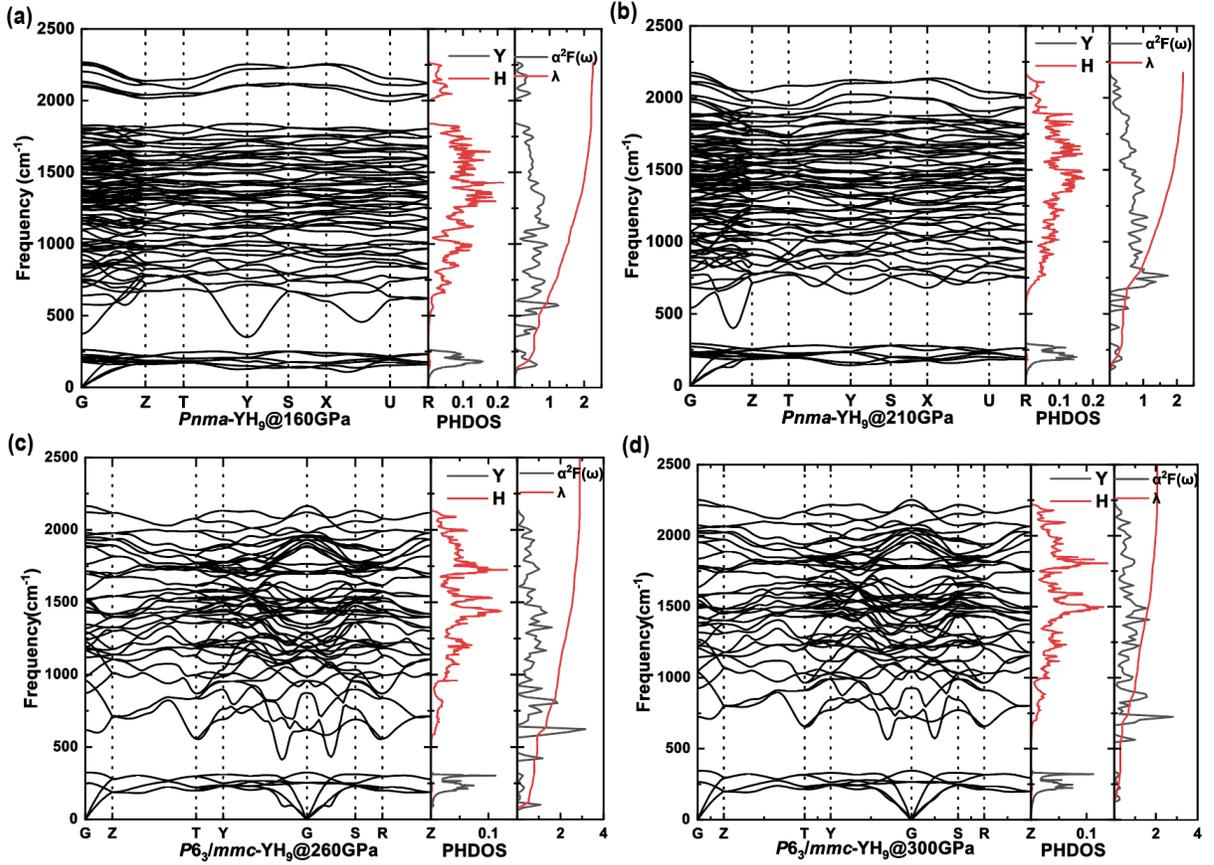

Fig. 7. Calculated phonon band structure, PHDOS, electron-phonon coupling (EPC) parameter λ, and Eliashberg spectral function $α^2F(ω)$ of electronic band structures and projected density of states for *Pnma* phase at (a) 160 GPa and (b) 210 GPa, $P6_3/mmc$-1 phase at (c) 260 GPa and (d) 300 GPa.

Although *Pnma* and $P6_3/mmc$-1 phases have similar structure, their pressure dependence of $T_c$ is different: with compression the $T_c$ of *Pnma* phase increases rapidly while the $T_c$ of $P6_3/mmc$-1 phase decreases. To further understand the different pressure dependence of $T_c$, we calculated the phonon band structure and Eliashberg spectral function $α^2F(ω)$ of *Pnma* and $P6_3/mmc$-1 phases at different pressures, as shown in Fig. 7. For *Pnma* phase in Fig. 7a and b, it can be seen from PHDOS that the low frequency phonon modes (below 250 cm$^{-1}$) are mainly from Y atom, medium frequency phonon modes (300-1800 cm$^{-1}$) are mainly from bending modes of H atom, and high frequency phonon modes (above 2000 cm$^{-1}$) are entirely from H$_2$ molecular units. As we discussed in Fig. 3, compression will promote a more even distribution of charges, thus making short bonds longer and long bonds shorter. With increasing pressure, short H-H bonds in H$_2$ molecular units stretch and high frequency vibration diminish, while long H-H bonds contract and medium frequency vibration enhance (see Fig. 7b), leading to the increase of average phonon frequency. Logarithmic average phonon frequency $\omega_{\log}$ of the *Pnma* phase increases from 890 K at 160 GPa to 1153 K at 210 GPa (see Table S3). We know that the increase of superconducting transition temperature mainly comes from the enhancement of electron-phonon coupling and the increase of average phonon frequency. For the *Pnma* phase, the $\lambda$ change a little with increase pressure, which is 2.26 at 160 GPa and 2.19 at 210 GPa. Therefore, the increase of $\omega_{\log}$ makes the $T_c$ of *Pnma* phase increases rapidly from 191-205 K to 233-250 K in the pressure range of 160-210 GPa.

For *P*6$_3$/*mmc*-1 phase, it exhibits prominent phonon "softening" at 260 GPa, and the softening becomes weaker with the increase of pressure, which leads to the weakening of electron-phonon coupling. The $\lambda$ is 2.90 at 250 GPa and 2.06 at 300 GPa. Although the $\omega_{\log}$ increases with compression from 790 K at 260 GPa to 1255 K at 300 GPa (see Table S3), it cannot counteract the influence of electron-phonon coupling weakening. Therefore, *P*6$_3$/*mmc*-1 phase shows a decreasing trend of $T_c$ with the increase of pressure, from 250-266 K to 234-252 K in the pressure range of 250-300 GPa.

**Conclusions**

In summary, we have systematically explored the crystal structures of YH$_9$ up to 300 GPa by random structure searching method, and found a lowly symmetric *Pnma* phase with distorted H$_{29}$ cage. This phase is predicted to be stable between 160 and 210 GPa, and its XRD and pressure dependence of $T_c$ are consistent with the experimental results. Further calculations show that the charge distribution in *Pnma* phase tends to be average with increasing pressure, which weaken the distortion of H$_{29}$ cage, and finally makes it exhibit a positive pressure dependence of $T_c$. This phenomenon has been experimentally established in LaH$_{10}$ and H$_3$S. Our findings not only provide a theoretical support to the experimental observations, but also shed light on the correlation between superconductivity and structural distortion in YH$_9$.

**Conflict of interest**

The authors declare no competing financial interest.

**ACKNOWLEDGMENT**

We are grateful to Vasily S. Minkov and Mikhail I. Erements who provide experimental data in Fig. 4. We are also grateful to Feng Peng who provide crystal structure of *Cmcm*-YH$_9$. This work was supported by the National Natural Science Foundation of China (Grants No. 12122405, No. 12274169 and No. 52072188), National Key R&D Program of China (Grant No. 2018YFA0305900 and No. 2022YFA1402304), and Jilin Provincial Science and Technology Development Project (20210509038RQ). Parts of calculations were performed in the High

**REFERENCES**

[1] E. Wigner and H. B. Huntington, J. Chem. Phys. **3**, 764 (1935).
[2] N. W. Ashcroft, Phys. Rev. Lett. **21**, 1748 (1968).
[3] R. P. Dias and I. F. Silvera, Science **355**, 715 (2017).
[4] P. Loubeyre, F. Occelli, and P. Dumas, Nature **577**, 631 (2020).
[5] N. W. Ashcroft, Phys. Rev. Lett. **92**, 4, 187002 (2004).
[6] D. F. Duan, Y. X. Liu, Y. B. Ma, Z. Shao, B. B. Liu, and T. Cui, Natl. Sci. Rev. **4**, 121 (2017).
[7] L. P. Gor'kov and V. Z. Kresin, Rev. Mod. Phys. **90**, 16, 011001 (2018).
[8] M. Du, W. Zhao, T. Cui, and D. Duan, Journal of Physics: Condensed Matter **34**, 173001 (2022).
[9] D. F. Duan *et al.*, Sci Rep **4**, 6, 6968 (2014).
[10] H. Y. Liu, Naumov, II, R. Hoffmann, N. W. Ashcroft, and R. J. Hemley, Proc. Natl. Acad. Sci. U. S. A. **114**, 6990 (2017).
[11] F. Peng, Y. Sun, C. J. Pickard, R. J. Needs, Q. Wu, and Y. M. Ma, Phys. Rev. Lett. **119**, 6, 107001 (2017).
[12] A. P. Drozdov, M. I. Eremets, I. A. Troyan, V. Ksenofontov, and S. I. Shylin, Nature **525**, 73 (2015).
[13] M. Einaga, M. Sakata, T. Ishikawa, K. Shimizu, M. I. Eremets, A. P. Drozdov, I. A. Troyan, N. Hirao, and Y. Ohishi, Nat. Phys. **12**, 835 (2016).
[14] A. P. Drozdov *et al.*, Nature **569**, 528 (2019).
[15] M. Somayazulu, M. Ahart, A. K. Mishra, Z. M. Geballe, M. Baldini, Y. Meng, V. V. Struzhkin, and R. J. Hemley, Phys. Rev. Lett. **122**, 6, 027001 (2019).
[16] H. Wang, J. S. Tse, K. Tanaka, T. Iitaka, and Y. Ma, Proc. Natl. Acad. Sci. U. S. A. **109**, 6463 (2012).
[17] X. Feng, J. Zhang, G. Gao, H. Liu, and H. Wang, RSC Adv. **5**, 59292 (2015).
[18] Y. Li, J. Hao, H. Liu, J. S. Tse, Y. Wang, and Y. Ma, Sci Rep **5**, 9948 (2015).
[19] K. Abe, Phys. Rev. B **96**, 144108 (2017).
[20] Hao Song, Zihan Zhang, Tian Cui, Chris J. Pickard, Vladimir Z. Kresin, and D. Duan, Chinese Physics Letters **38**, 107401 (2021).
[21] H. Xie *et al.*, J. Phys.-Condes. Matter **31**, 7, 245404 (2019).
[22] X. Liang, A. Bergara, L. Wang, B. Wen, Z. Zhao, X.-F. Zhou, J. He, G. Gao, and Y. Tian, Phys. Rev. B **99**, 100505 (2019).
[23] W. Zhao, D. Duan, M. Du, X. Yao, Z. Huo, Q. Jiang, and T. Cui, Phys. Rev. B **106**, 014521 (2022).
[24] W. Sukmas, P. Tsuppayakorn-aek, U. Pinsook, and T. Bovornratanaraks, J. Alloy. Compd. **849**, 156434 (2020).
[25] L.-T. Shi, Y.-K. Wei, A. K. Liang, R. Turnbull, C. Cheng, X.-R. Chen, and G.-F. Ji, Journal of Materials Chemistry C **9**, 7284 (2021).
[26] P. Song, Z. Hou, P. B. d. Castro, K. Nakano, K. Hongo, Y. Takano, and R. Maezono, Chem. Mat. **33**, 9501 (2021).
[27] M. Du, H. Song, Z. Zhang, D. Duan, and T. Cui, Research **2022**, 9784309 (2022).
[28] D. V. Semenok *et al.*, Mater. Today **33**, 36 (2020).
[29] P. Kong *et al.*, Nat. Commun. **12**, 5075 (2021).
[30] E. Snider, N. Dasenbrock-Gammon, R. McBride, X. Wang, N. Meyers, K. V. Lawler, E. Zurek, A. Salamat, and R. P. Dias, Phys. Rev. Lett. **126**, 117003 (2021).
[31] Y. Wang, K. Wang, Y. Sun, L. Ma, Y. Wang, B. Zou, G. Liu, M. Zhou, and H. Wang, Chinese Physics B (2022).
[32] L. Ma *et al.*, Phys. Rev. Lett. **128**, 167001 (2022).
[33] Z. Li *et al.*, Nat. Commun. **13**, 2863 (2022).
[34] I. A. Troyan *et al.*, Adv. Mater. **33**, 2006832 (2021).
[35] W. Chen, D. V. Semenok, X. Huang, H. Shu, X. Li, D. Duan, T. Cui, and A. R. Oganov, Phys. Rev. Lett. **127**, 117001 (2021).
[36] D. V. Semenok *et al.*, Mater. Today **48**, 18 (2021).

[37] W. Chen, X. Huang, D. V. Semenok, S. Chen, K. Zhang, A. R. Oganov, and T. Cui, 2022), p. arXiv:2203.14353.
[38] J. K. Bi *et al.*, Nat. Commun. **13**, 5952 (2022).
[39] M. Shao, W. Chen, K. Zhang, X. Huang, and T. Cui, Phys. Rev. B **104**, 174509 (2021).
[40] D. Y. Kim, R. H. Scheicher, and R. Ahuja, Phys. Rev. Lett. **103**, 077002 (2009).
[41] X. Huang *et al.*, Natl. Sci. Rev. **6**, 713 (2019).
[42] D. Sun *et al.*, Nat. Commun. **12**, 6863 (2021).
[43] A. M. Shipley, M. J. Hutcheon, R. J. Needs, and C. J. Pickard, Phys. Rev. B **104**, 054501 (2021).
[44] C. J. Pickard and R. J. Needs, Phys. Rev. Lett. **97**, 4, 045504 (2006).
[45] C. J. Pickard and R. J. Needs, J. Phys.-Condes. Matter **23**, 23, 053201 (2011).
[46] S. J. Clark, M. D. Segall, C. J. Pickard, P. J. Hasnip, M. J. Probert, K. Refson, and M. C. Payne, Z. Kristall. **220**, 567 (2005).
[47] J. P. Perdew, K. Burke, and M. Ernzerhof, Phys. Rev. Lett. **77**, 3865 (1996).
[48] Kresse and Furthmuller, Phys. Rev. B **54**, 11169 (1996).
[49] D. J. Chadi, Phys. Rev. B **16**, 1746 (1977).
[50] G. Kresse and D. Joubert, Phys. Rev. B **59**, 1758 (1999).
[51] P. Giannozzi *et al.*, J. Phys.-Condes. Matter **21**, 19, 395502 (2009).
[52] W. L. McMillan, Phys. Rev. **167**, 331 (1968).
[53] P. B. Allen and R. C. Dynes, Phys. Rev. B **12**, 905 (1975).
[54] G. M. Eliashberg, Sov Phys Jetp **11:3**, 696 (1960).
[55] D. Zhou *et al.*, Science Advances **6**, eaax6849 (2020).